\documentclass[letterpaper, 10pt, conference]{IEEEconf}
\IEEEoverridecommandlockouts 

\usepackage{include}

\begin{document}

\title{\LARGE \bf Considerate and Cooperative Model Predictive Control for Energy-Efficient Truck Platooning of Heterogeneous Fleets}

\author{
    {Tyler Ard,~ Bibin Pattel,~ Ardalan Vahidi,~ Hoseinali Borhan}%
    \thanks{Tyler Ard ({\tt\footnotesize trard@g.clemson.edu}) and Ardalan Vahidi ({\tt\footnotesize avahidi@g.clemson.edu}) are with the Department of Mechanical Engineering, Clemson University, Clemson, SC 29634, USA.}%
    \thanks{Bibin Pattel ({\tt\footnotesize bibin.n.pattel@cummins.com}) and Hoseinali Borhan ({\tt\footnotesize hoseinali.borhan@cummins.com}) are with Cummins Inc, Columbus, IN 47201, USA.}%
}


\maketitle

\begin{abstract}
Connectivity-enabled automation of distributed control systems allow for better anticipation of system disturbances and better prediction of the effects of actuator limitations on individual agents when incorporating a model. 
Automated convoy of heavy-duty trucks in the form of platooning is one such application designed to maintain close gaps between trucks to exploit drafting benefits and improve fuel economy, and has traditionally been handled with classically-designed connected and adaptive cruise control (CACC). This paper is motivated by demonstrated limitations of such a control strategy, in which a classical CACC was unable to efficiently handle real-world road grade and velocity transient disturbances without the assistance of fleet operator intervention, and is non-adaptive to varied hardware and loading conditions of the operating truck. This automation strategy is addressed by forming a cooperative model predictive control (MPC) for eco-platooning that considers interactions with trailing trucks to incentivize platoon harmonization under road disturbances, velocity transients, and engine limitations, and further improves energy economy by reducing unnecessary engine effort. This is accomplished for each truck by sharing load, maximum engine power, transmission ratios, control states, and intended trajectories with its nearest neighbors. The performance of the considerate and cooperative strategy was demonstrated on a real-world driving scenario against a similar non-considerate control strategy, and overall it was found that the considerate strategy significantly improved harmonization between the platooned trucks in a real-time implementable manner.
\end{abstract}


%

\section{Introduction}

CACC approaches have been shown effective in passenger vehicles to both improve traffic capacity and road efficiency \cite{Shladover2012}, and to improve fuel economy through system-wide traffic harmonization effects and reduction of braking \cite{Liu2020}. Overall, network connectivity has been shown essential to boost cooperation of automated agents on the road and promote positive vehicle-to-vehicle (V2V) interactions for a mutual benefit \cite{Wang2020}.

\begin{figure}
    \centering
    \vspace{1em}
    \input{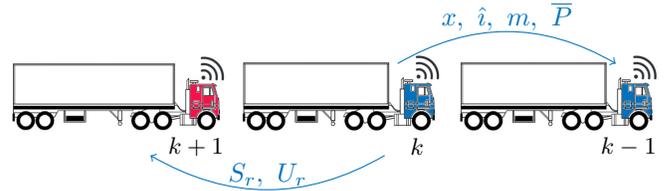}
    \caption{Bi-directional connected and cooperative heterogeneous truck platoon.}
    \label{fig:Topology}
\end{figure}

Specifically in the context of truck platooning - in which heavy-duty trucks are driven in tight formations to exploit drafting benefits and improve fuel economy from reduced aerodynamic drag and consequently engine demand - there is a need for introducing advanced motion planning strategies to consistently achieve robust and optimal performance. A classical CACC controller is experimented on in \cite{McAuliffe2018} for a 3-truck platoon and is demonstrably effective in reducing fuel consumption under a closed-loop test track. However, experimental performance of classical CACC platooning under real-world conditions is further studied in \cite{Borhan2021} when tasked to compensate for real-world driving conditions including road grade and traffic cycles. The real-world test results in this study expose some problematic performance in traditional CACC approaches when under heavy disturbances that are not properly compensated, and can completely remove the benefits of platooning a truck convoy. By this, under typical real-world driving conditions with traffic, heavy road grade, and heterogeneous loading conditions, platooning systems require advanced motion planning techniques over traditional methods to compensate. Figure \ref{fig:Topology} displays trucks of heterogeneous makeup arranged into a connected platooning configuration. 

Look-ahead predictive control, in particular, is a desirable choice for motion planning strategies to optimally guide a system subject to explicit performance criteria and constraints. Actuation limitations and dynamics of a system are accounted for using a model, and ultimately influence the set of states that are reachable in a finite amount of time when possibly subject to counter-acting disturbances \cite{Kerrigan2000}. Interactions against other agents in particular rely on anticipating their motion, and rely on an educated prediction unless direct communications are available. Distributed MPC can form a joint optimization problem between the ego agent and other agents, such as the approach proposed in \cite{Keviczky2006}, that considers a model of surrounding agents directly in the ego agent optimization to aid in prediction. Other agents are similarly considered in \cite{Dunbar2006} during optimization for a vehicle platoon, and their control actions are constrained to follow a group-expected behavior. For vehicle platoon formation and motion planning, a one-way communication strategy is shown in \cite{Zheng2017} to stabilize.

Anticipative motion planning in automated driving has already been shown to significantly boost performance. Eco-platooning approaches have been previously proposed in model-based frameworks utilizing a communication network for guaranteed collision-avoidance in passenger vehicles \cite{Wang2015}, for improved energy economy by cutting unnecessary braking \cite{Turri2017}, and to exploit drafting effects for reduced coefficient of drag in heavy trucks \cite{Turri2017b}. Vehicle awareness can be sourced from manually-driven vehicles for passive connectivity as well, and has been shown to further aid in predicting upcoming traffic disturbances \cite{Ibrahim2019}. In the motion planning of automated trucks for improved fuel economy, fuel-optimal trajectories are analytically derived in \cite{He2016} to guide a heavy truck through a road profile and nominal traffic disturbance, whereas \cite{Hellstrom2009} implement a numerical fuel-optimal strategy for the motion of a heavy truck during experimentation. 



\section{Problem Motivation}\label{sec:Motivation}

\begin{figure}
    \centering
    \input{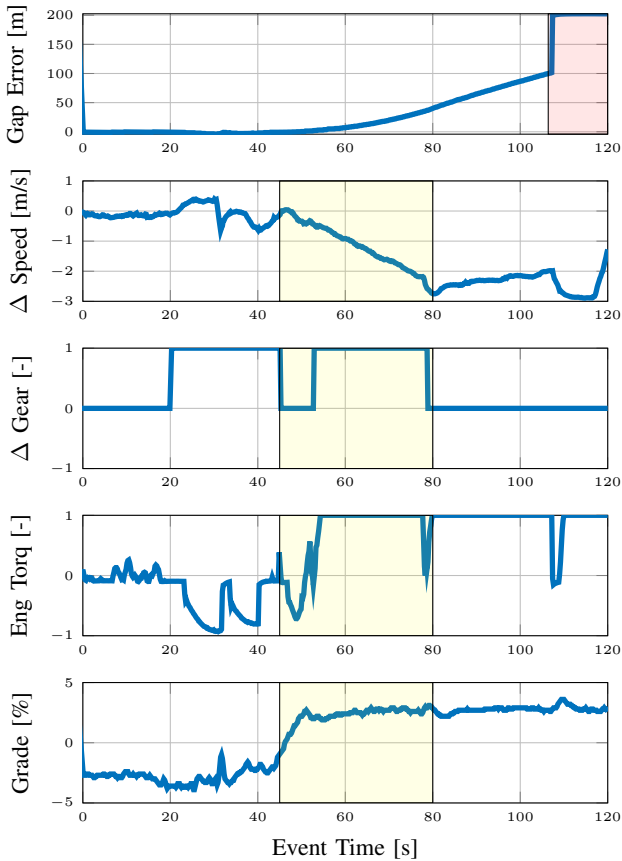}
    \caption{Experimental test data excerpted from a CACC-enabled platooned truck in a fleet \cite{Borhan2021}. The highlighted yellow region indicates a change to uphill operation where an up-shifted gear leads to reduced ego climbing ability and increased gap error; and eventually disengagement of the CACC - indicated by the red region and discontinuous jump in gap signal. Operator $\Delta(\cdot)$ indicates difference in ego truck quantity subtracted by preceding truck quantity.}
    \label{fig:Motivation}
\end{figure}

This paper is motivated by recent experimental results observed on the highway when engaging a classically-designed CACC for platooning trucks \cite{Borhan2021}. In this, several freight trucks of same mass and hardware specifications were deployed in a platooning configuration to travel a route with expected road grade and traffic conditions for heavy-duty long-haul truck applications. The CACC was designed to closely track desired intervehicle gaps to enable safe operation (see, for example, \cite{McAuliffe2018}, \cite{Ploeg2011}). However, Figure \ref{fig:Motivation} illustrates a data excerpt exposing some problematic control performance of the CACC when under transients in preceding truck speed, and when under a disturbance combination of 1) heavy grade and 2) an up-shifted gear from the automatic transmission compared to the preceding truck. The up-shifted gear reduced the ratio of torque applied at the wheel and thus reduced its grade-climbing and acceleration capabilities. Ultimately, despite maximizing allowable engine torque, the ego truck cannot maintain its platoon with its leader, and the CACC becomes disengaged. Furthermore, the platoon can be difficult and inefficient to re-form by expending fuel to accelerate and close the gaps between trucks, or when in the presence of surrounding traffic due to allowing cut-ins that need to be maneuvered around.

The scenario shown in Figure \ref{fig:Motivation} is a difficult, albeit not uncommon, one when managing a fleet. In it, the ego truck maximized its engine effort and was not able to keep up with the preceding truck. In fact, the demonstrated scenario was a special case in which the deployed trucks in the fleet had the same mass and hardware specifications, and therefore same propulsive capabilities, but were still subject to gap-tracking issues in highway conditions. In sight of this, we propose that with autonomy-enabled platooning, a more cooperative strategy is needed from leading trucks to assist in maintaining platoon compactness and stability, which we refer to as \emph{consideration}.

This paper proposes the design of a model-based predictive controller for the longitudinal motion planning required in real-world roads with traffic and road grade. To compensate for propulsive differences that can arise from a variety of conditions including 1) asynchronous gearing, 2) heterogeneous loading conditions, and 3) differing powertrain capabilities, a cooperative and distributed model predictive controller is proposed that solves a joint optimization problem considering itself and its trailing truck. In doing so, the ego truck selects actions for itself to be applied in a receding horizon fashion, as well as aids its own prediction by computing suggested actions for its trailing truck that are communicated but not applied. 

\section{Platoon Modeling}

A longitudinal simulation model of the platoon dynamics is developed and verified against empirically gathered data considering 1) powertrain slew rate constraints that limit abrupt changes in input power, 2) engine torque conversion through a gearbox to applied force at the wheel, 3) the slipstream effects that reduce aerodynamic friction when following behind a tractor-trailer truck, and 4) fuel consumption based on engine torque $\tau$ and engine speed $\omega$ \cite{GUZZELLA2013}.

Unifying force at the wheel kinetics are used to predict the longitudinal response of the truck 
\begin{equation}
    m_e(\hat{\imath}) \dot{v} = F - F_a(d) - F_r(s)
\end{equation}
where $F = ma_t$ is the applied tractive force at the wheel, $m$ is the total truck mass, $a_t$ is applied tractive acceleration, and $m_e(\hat{\imath})$ is effective inertial mass of the truck considering rotating components with the current gear $\hat{\imath}$.

\begin{figure}
    \centering
%
%
\definecolor{mycolor1}{rgb}{0.00000,0.44700,0.74100}%
\begin{tikzpicture}

\begin{axis}[%
width=0.808\columnwidth,
height=0.517\columnwidth,
at={(0\columnwidth,0\columnwidth)},
scale only axis,
xmin=10.00000,
xmax=110.00000,
xlabel style={font=\color{white!15!black}},
xlabel={Gap (d) [m]},
ymin=0.88000,
ymax=1.00000,
ytick={0.88, 0.91, 0.94, 0.97, 1.0},
ylabel style={font=\color{white!15!black}},
ylabel={Drag Reduction [\%]},
axis background/.style={fill=white},
xmajorgrids,
ymajorgrids,
legend style={at={(0.97,0.03)}, anchor=south east, legend cell align=left, align=left, draw=white!15!black},
ylabel near ticks,
xlabel near ticks,
ticklabel style={font=\small}
]
\addplot[only marks, mark=*, mark options={}, mark size=1.5000pt, color=black, fill=black] table[row sep=crcr]{%
x	y\\
15.00000	0.90497\\
20.00000	0.91298\\
30.00000	0.92834\\
40.00000	0.93729\\
50.00000	0.94624\\
60.00000	0.95519\\
70.00000	0.96415\\
80.00000	0.97310\\
100.00000	0.99100\\
};
\addlegendentry{Data}

\addplot [color=mycolor1, line width=2.0pt]
  table[row sep=crcr]{%
10.00000	0.89314\\
11.00000	0.89576\\
12.00000	0.89821\\
13.00000	0.90052\\
14.00000	0.90270\\
15.00000	0.90476\\
16.00000	0.90671\\
17.00000	0.90856\\
18.00000	0.91033\\
19.00000	0.91201\\
20.00000	0.91362\\
21.00000	0.91515\\
22.00000	0.91663\\
23.00000	0.91805\\
24.00000	0.91942\\
26.00000	0.92203\\
28.00000	0.92448\\
30.00000	0.92680\\
32.00000	0.92902\\
34.00000	0.93115\\
36.00000	0.93321\\
39.00000	0.93619\\
42.00000	0.93908\\
46.00000	0.94282\\
51.00000	0.94737\\
57.00000	0.95273\\
68.00000	0.96246\\
85.00000	0.97750\\
96.00000	0.98732\\
107.00000	0.99723\\
110.00000	0.99995\\
};
\addlegendentry{$\beta(d) = A\exp(Bd) + C\exp(Dd)$}

\end{axis}
\end{tikzpicture}%
    \caption{Platooning drag reduction benefit by inter-vehicle gaps experimentally identified under test track conditions.}
    \label{fig:DragReduction}
\end{figure}
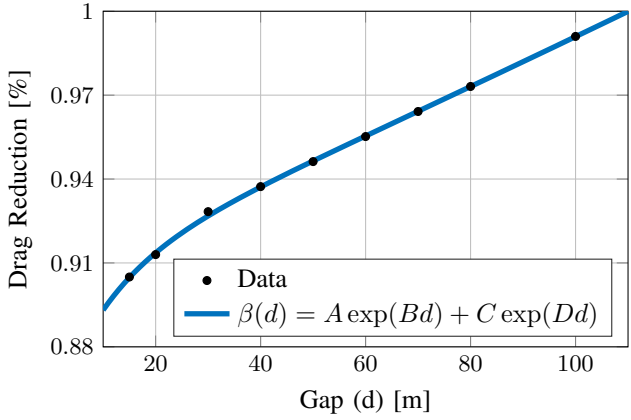

The longitudinal aerodynamic drag force $F_a$ when subject to no external wind is expressed as a function of vehicle speed $v$ and possibly a function of inter-vehicle gap $d$ if following behind another truck:
\begin{equation}
    F_a(d) = \displaystyle\begin{cases} \frac{1}{2}\rho A_f C_D \beta(d) v^2 & d \in [0, 110] \text{ following a truck} \\ \frac{1}{2}\rho A_f C_D v^2 & \text{otherwise} \end{cases}
\end{equation}
where $\rho$ is the density of air, $A_f$ is the effective frontal surface area of the truck, $C_D$ is the nominal drag coefficient, and $\beta(d)$ is a drag reduction function. This drag reduction is empirically available and can be expressed as in Figure \ref{fig:DragReduction}.

Similarly, the longitudinal rolling forces $F_r$ are expressed as a function of road grade $\alpha$ that varies with position $s$:
\begin{equation}
    F_r(s) = mg\left( C_r \cos{\alpha(s)} + \sin{\alpha(s)} \right)
\end{equation}
where $g$ is the gravitational constant and $C_r$ is the rolling resistance. Sparse preview of upcoming road grade is readily available from infrastructure-connected technology, and to treat this, a Legendre polynomial series is fit in a least-squares fashion to develop a form for the road grade.
\begin{equation}
    \alpha(\ell) \approx \sum_{j=0}^3 c_j P_j(\ell), \ \ell \in [-1, 1]
\end{equation}

Here, $\ell$ is a linearly-mapped variable of the position interval available within a preview horizon $s \in [0, s_+] \mapsto [-1, 1]$, $c$ is a fitting coefficient, and $P_j$ is the Legendre polynomial of the $j$-th kind.

Vehicle propulsion through wheel force $F$ is proportionally related to applied engine torque $\tau$ by a gearbox reduction 
\begin{equation*}
    F = \frac{\tau_w}{r_w} = \tau \frac{\eta(\hat{\imath})  \hat{\imath}_f \hat{\imath}_r(\hat{\imath})}{r_w}
\end{equation*}
where $\tau_w$ is the torque at the wheel, $r_w$ is the wheel radius, $\hat{\imath}$ is the current gear, $\eta$ is the conversion efficiency, $\hat{\imath}_f$ is the final drive ratio, and $\hat{\imath}_r$ is the current gear ratio.

It follows that maximum force at the wheel from the engine $\overline{F}$ is expressed by the maximum engine torque $\overline{\tau}$
\begin{equation}\label{eq:force}
    |F| \leq \overline{F}(\hat{\imath}) = \overline{\tau} \ \frac{\eta(\hat{\imath}) \hat{\imath}_f \hat{\imath}_r(\hat{\imath})}{r_w}
\end{equation}
and also, the force at the wheel is limited by the maximum engine power $\overline{P}$ in an isometric, non-convex fashion
\begin{equation}\label{eq:isometric}
    F v \leq \overline{P}
\end{equation}

Finally, engine torque slew is limited by a possibly discontinuous function $$ \dot{\tau} \leq g(\omega, \tau), $$ and for the purposes of control modeling can be approximated by a lag equation. Choosing a first-order lag equation with time-constant $\tau_d$, the tractive acceleration command $u$ affects the currently applied tractive acceleration $a_t$ by
\begin{equation}
    \tau_d\dot{a}_t + a_t = u
\end{equation}

The state vector is then expressed as $$ x(t) = \begin{bmatrix} s(t) & v(t) & a_t(t) \end{bmatrix}^\intercal $$
and so the state-space model follows as 
\begin{equation}
\begin{split}
    \dot{x} &= f(x, u, w) \\
    &= \begin{bmatrix}
    v \\
     \displaystyle\frac{1}{m_e(\hat{\imath})} \Big(ma_t - F_a(d) - F_r\left(s\right)\Big) \\
     \tau_d^{-1}\left( u-a_t \right)
    \end{bmatrix}
\end{split}
\end{equation}
with the admissible control set $\mathcal{U}$ formed by the relations \eqref{eq:force} and \eqref{eq:isometric}.


\section{Considerate Model Predictive Control}

Section \ref{sec:Motivation} identified challenges in automation and deployment for truck platooning on real roads when using traditional classes of control. In particular, transient operation under highway conditions was identified to lead to dissolution of the platoon - stemming from heavy road grade and heterogeneous loading and hardware conditions \cite{Borhan2021}. 

In response to these experimentally observed challenges in truck platooning in real-world driving conditions, further considerate motion planning strategies are proposed to compensate and assist less-capable trucks. The goals of such strategies are to take actions to 1) maintain platoonable gaps during highway operation, 2) improve velocity synchronization between platooned trucks, 3) reduce disengagements of automated longitudinal control to reduce driver intervention, and 4) improve fuel economy of the total platoon. 

The strategy relies on a bi-directional communication topology as illustrated in Figure \ref{fig:Topology} - in which truck $k$ receives necessary control parameters from truck $k+1$ behind it: current gear $\hat{\imath}^{(k+1)}$; current gap, velocity, and traction for state $x^{(k+1)}$; mass $m^{(k+1)}$; and engine power and torque limitations $\overline{P}^{(k+1)}$. Truck $k$ solves a distributed motion planning problem for itself that considers the truck behind it, in which it takes actions to: regulate their combined control efforts, help truck $k+1$ maintain its desired gap, and, if it is the leader of the platoon, track a target desired velocity. Solving the combined motion planning problem aids in predicting the response the ego can take that benefits both trucks. Truck $k$ then broadcasts its planned forward positions $S_r$ and suggested control actions $U_r$ for truck $k+1$ to follow as a reference for its own motion planning, which serve as a soft level of compliance so that each following truck is behaving as expected by others in the platoon \cite{Dunbar2006}.

The cost for a truck $k$ is expressed as the following, where superscript parenthetic terms indicate logic for position-in-the-platoon-dependent quantities (a term \scriptsize$(k|k=1)$\normalsize \ applies if an ego truck is the leader of the platoon, for example) 
\begin{equation*}
\begin{split}
    J^{(k)} &= q_t \left(\frac{s_f-s_N^{(k)}}{t_f-t_N}-v_N^{(k)}\right)^2 + 
    \sum_{i=0}^{N-1}\bigg[ 
    q_u \left(u_i^{(k)}\right)^2 \\ &
    + q_v\left((v_i-\nu)^{(k|k=1)}\right)^2 + q_d\left( (d_i - Tv_i)^{(k|k>1)} \right)^2
    \\ & 
    + q_{c}\left( (u_i - \mu_i)^{(k|k>1)} \right)^2 \bigg] + q_\epsilon \epsilon^{(k)}
\end{split}
\end{equation*}

Here, $i$ indicates the stage of the control problem with $N$ total stages, $\epsilon$ is a slack decision variable used to soften the gap constraint, $\mu_i \in U_r$ is the stage-dependent control action as suggested from the preceding truck, $\nu$ is the in-horizon velocity reference, $d \triangleq s^{(k-1)}-L^{(k-1)}-s^{(k)}$ is the intervehicle gap considering the length $L$ of the preceding truck and with $s^{(k-1)} \in S_r$ available from vehicle connectivity, and $T$ is a desired following headway time. A reference terminal speed is added that tracks an average velocity needed to reach the remaining distance-to-go in the trip $s_f-s_N$ in the remaining desired time-to-go $t_f - t_N$ at the end of the horizon. Variables $q$ indicate weightings between each term in the performance metric.

State constraints are additionally imposed to handle safety and enforce road laws. The velocity is limited to prevent reversing and excessive speeding, and gap is constrained to avoid getting too close to the preceding vehicle.
\begin{equation} \label{eq:constraints}
    %
    \begin{split}
        0 &\leq v_i^{(k)} + \epsilon_1^{(k)} \leq \overline{v} \\
        \underline{d} &\leq d_i^{(k)} + \epsilon_2^{(k)}
    \end{split}
\end{equation}
Equations \eqref{eq:constraints} then form the admissible states $\mathcal{X}$ for truck $k$.

In summary, each truck $k \in 1, \ldots, K-1$ solves a \emph{considerate} motion planning problem as in the following

\begin{mini*}|l|
    { U^{(k)}, U^{(k+1)} }
    {J^{(k)} + J^{(k+1)}
    }{}{}
    \addConstraint{
        \dot{x}^{(k)}}{ = f\left(x^{(k)}, u^{(k)}, w^{(k)}\right)
    }{}
    \addConstraint{
        \dot{x}^{(k+1)}}{ = f\left(x^{(k)}, x^{(k+1)}, u^{(k+1)}, w^{(k+1)}\right)
    }{}
    \addConstraint{
        u^{(k)} \in \mathcal{U}^{(k)}, 
    }{\ x^{(k)} \in \mathcal{X}^{(k)}}
    \addConstraint{u^{(k+1)} \in \mathcal{U}^{(k+1)},}{\ x^{(k+1)} \in \mathcal{X}^{(k+1)}
    }{}
\end{mini*}

\noindent applying the first action in $U^{(k)}$ only. Similarly, the final truck in the platoon $k = K$ solves an optimization for itself.

\begin{mini*}|l|
    { U^{(k)} }
    {J^{(k)}
    }{\label{eq:nonCooperative}}{}
    \addConstraint{
        \dot{x}^{(k)}}{ = f\left(x^{(k)}, u^{(k)}, w^{(k)}\right)
    }{}
    \addConstraint{
        u^{(k)} \in \mathcal{U}^{(k)},
    }{\ x^{(k)} \in \mathcal{X}^{(k)}}
\end{mini*}

Finally, an \emph{anticipative} variant of the control strategy is designated for trucks $k \in 1, \ldots, K$ in which the previous optimization is solved with additionally the compliance weight $q_c = 0$. This control strategy has been previously studied in \cite{ARD2020} against a flat road, and is already a significant improvement in fuel performance over a non-anticipative strategy due to reduced engine demand.

Each optimization results in a non-convex non-linear program that must be solved to an optima local to a supplied initial guess. The \texttt{Forces Pro} solver is used with \texttt{Casadi} to generate efficient \texttt{C} code to solve the problem 
\cite{FORCESNLP}, \cite{Andersson2019}. For the considerate optimization with $N=22$ and 10 inter-stage integration nodes, the problem solves in \unit[25]{ms} on average with a maximum solution time of \unit[60]{ms} on a 4-core \unit[2.8]{GHz} processor. The anticipative strategy solved in \unit[12]{ms} on average with a maximum solution time of \unit[40]{ms}.

\section{Results}

A 3-truck platoon of varied loading conditions is constructed and simulated to demonstrate controller performance. The considerate controller (C) and anticipative controller (A) variants are both first demonstrated on a nominal S-Road shaped profile of a heavy downhill and heavy uphill section of road, where the leading truck is at an unloaded mass of \unit[14]{t} and the two trailing trucks are at a heavy loaded mass of \unit[38]{t}.

\begin{figure}
    \centering
%
%
\newcommand{\len}{2.0}
\definecolor{mycolor1}{rgb}{0.00000,0.44700,0.74100}%
\definecolor{mycolor2}{rgb}{0.85000,0.32500,0.09800}%
\definecolor{mycolor3}{rgb}{0.92900,0.69400,0.12500}%
\begin{tikzpicture}[%
spy using outlines={circle,lens={scale=1.7}, size=3cm, connect spies}
]
\begin{axis}[%
width=0.829\columnwidth,
height = 0.437\columnwidth,
at={(0\columnwidth,0\columnwidth)},
scale only axis,
xmin=-0.0360,
xmax=66.5321,
xlabel style={font=\color{white!15!black}},
xlabel={Position [km]},
ymin=0.0000,
ymax=2900.0000,
ylabel style={font=\color{white!15!black}},
ylabel={Time [s]},
ytick={0,500,1000,1500,2000,2500},
axis background/.style={fill=white},
xmajorgrids,
ymajorgrids,
legend style={at={(0.97,0.03)}, anchor=south east, legend cell align=left, align=left, draw=white!15!black},
ylabel near ticks,
xlabel near ticks,
ylabel style={font=\small},
xlabel style={font=\small},
legend style={font=\small},,
ticklabel style={font=\tiny}
]
\addplot [color=mycolor1, line width=\len pt, postaction={
        decoration={
          markings,
          mark=between positions 0.02 and 1 step 0.04
               with { \fill circle[radius=1.3pt]; },
        },
        decorate,
      },]
  table[row sep=crcr]{%
-0.0120	0.0000\\
0.0069	80.6000\\
0.0641	86.8000\\
0.1911	96.4000\\
0.3457	105.2000\\
0.5867	116.6000\\
0.8788	128.6000\\
1.2922	143.8000\\
2.2012	175.6000\\
5.3211	282.8000\\
23.5121	908.8000\\
29.2826	1108.2000\\
30.8385	1163.6000\\
32.4344	1221.8000\\
34.6915	1305.6000\\
35.5208	1337.6000\\
36.9103	1393.6000\\
37.6488	1424.6000\\
38.2023	1449.4000\\
38.7590	1476.2000\\
40.0722	1540.6000\\
40.6926	1573.2000\\
42.2303	1659.6000\\
43.3494	1719.2000\\
47.0091	1911.2000\\
51.8257	2162.6000\\
52.4614	2193.6000\\
53.1200	2223.8000\\
55.2029	2315.8000\\
55.8348	2341.4000\\
56.4503	2364.8000\\
58.4385	2438.0000\\
60.0507	2496.0000\\
66.3825	2720.0000\\
66.4588	2725.0000\\
66.5051	2730.4000\\
66.5278	2738.0000\\
66.5321	2760.4000\\
66.5321	2900.0000\\
};
\addlegendentry{\scriptsize C: k = 1}

\addplot [color=mycolor2, line width=1.0 pt, postaction={
        decoration={
          markings,
          mark=between positions 0 and 1 step 0.04
               with { \fill circle[radius=1.5pt]; },
        },
        decorate,
      },]
  table[row sep=crcr]{%
-0.0360	0.0000\\
-0.0146	85.0000\\
0.0642	92.6000\\
0.1974	102.2000\\
0.3657	111.6000\\
0.5915	122.2000\\
0.8833	134.2000\\
1.2809	148.8000\\
2.1778	180.2000\\
29.4135	1118.2000\\
30.9668	1173.6000\\
32.6651	1235.6000\\
34.8145	1315.4000\\
35.5819	1345.2000\\
37.4902	1423.0000\\
38.0748	1448.8000\\
38.5583	1471.8000\\
40.2360	1554.4000\\
40.8569	1587.6000\\
41.9288	1648.6000\\
42.7510	1693.0000\\
44.6349	1792.2000\\
51.6919	2161.2000\\
52.3452	2193.4000\\
52.9932	2223.4000\\
54.8350	2305.4000\\
55.5048	2333.4000\\
56.1463	2358.6000\\
57.0736	2393.2000\\
58.8970	2460.0000\\
60.5421	2518.8000\\
66.3018	2721.8000\\
66.4039	2726.6000\\
66.4603	2731.0000\\
66.4967	2737.0000\\
66.5106	2746.4000\\
66.5119	2900.0000\\
};
\addlegendentry{\scriptsize C: k = 3}

\addplot [color=mycolor3, dashed, line width=\len pt]
  table[row sep=crcr]{%
-0.0120	0.0000\\
0.0065	38.6000\\
0.0600	46.6000\\
0.1564	55.2000\\
0.2910	64.0000\\
0.4769	73.4000\\
0.7342	84.4000\\
1.0342	95.6000\\
1.7951	122.0000\\
32.0573	1164.2000\\
35.6683	1290.0000\\
36.4372	1318.2000\\
37.0434	1341.8000\\
38.9011	1417.8000\\
39.5033	1444.4000\\
40.7598	1503.8000\\
42.1835	1568.8000\\
51.7502	2001.4000\\
53.3114	2074.8000\\
53.9478	2102.6000\\
54.5348	2126.6000\\
55.1058	2148.4000\\
56.1825	2187.4000\\
57.1702	2222.0000\\
61.4448	2371.0000\\
61.8885	2389.0000\\
62.6088	2417.2000\\
63.0413	2435.6000\\
64.8403	2505.0000\\
66.3849	2564.8000\\
66.4592	2570.0000\\
66.5081	2576.2000\\
66.5297	2584.2000\\
66.5321	2606.6000\\
66.5321	2900.0000\\
};
\addlegendentry{\scriptsize A: k = 1}

\addplot [color=black!90, dashed, line width=\len pt]
  table[row sep=crcr]{%
-0.0360	0.0000\\
-0.0155	85.0000\\
0.0570	92.0000\\
0.1911	101.6000\\
0.3620	111.0000\\
0.5907	121.6000\\
0.8861	133.6000\\
1.2462	146.6000\\
35.0325	1310.0000\\
35.5958	1331.0000\\
36.1668	1353.6000\\
37.1553	1394.4000\\
37.7765	1421.6000\\
38.2693	1444.8000\\
40.2075	1540.2000\\
40.7873	1571.6000\\
41.5117	1613.2000\\
42.2675	1654.6000\\
43.9169	1742.4000\\
50.8893	2110.6000\\
52.6050	2197.8000\\
53.2294	2226.4000\\
55.2629	2316.0000\\
55.9033	2342.2000\\
56.4981	2365.0000\\
57.3590	2396.2000\\
58.1676	2424.4000\\
66.4897	2711.4000\\
66.5124	2714.0000\\
66.5134	2900.0000\\
};
\addlegendentry{\scriptsize A: k = 3}

\coordinate (spypoint) at (axis cs:45,1750);
\coordinate (magnifyglass) at (1.75cm,3.3cm);
  
\end{axis}

\spy [black, size=3.0cm] on (spypoint)
   in node[fill=white] at (magnifyglass);

\end{tikzpicture}%
    \caption{First and last trucks of the heterogeneous platoon travelling through the S-Road profile. The considerate (C) MPC maintains the platoon, whereas the anticipative (A) MPC does not maintain the platoon.}
    \label{fig:Travel}
\end{figure}
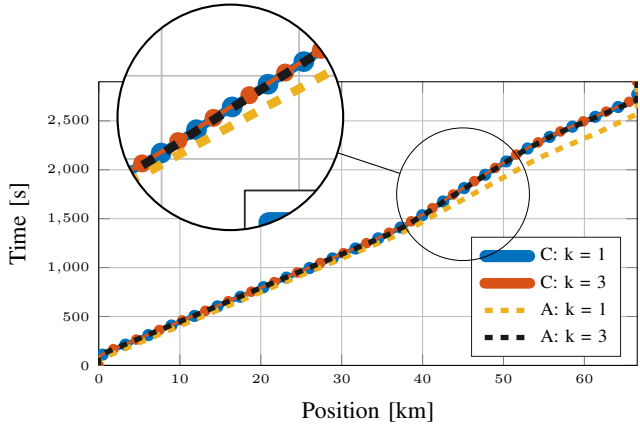

Figure \ref{fig:Travel} indicates time taken to reach a given position along the road for the first and last trucks in the platoon for each controller. Although the anticipative leading truck finishes the route sooner, it breaks away from the heavier, slower trucks and dissolves the platoon. However, the considerate strategy maintained the platoon for the entire route and finished in a similar time to the trailing trucks of the A strategy.

\begin{figure}
    \centering
    \input{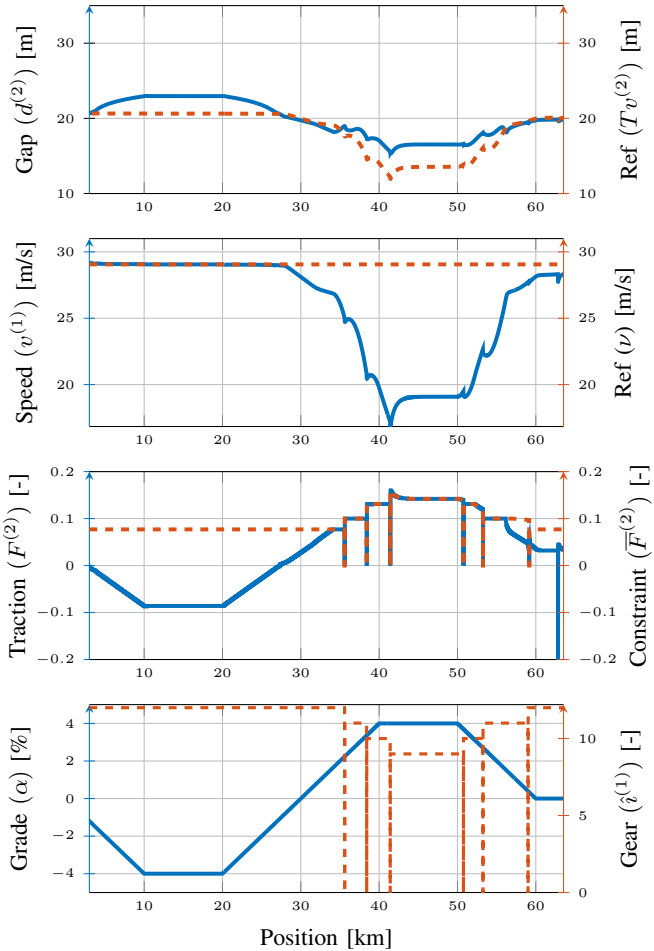}
    \caption{Performance of the considerate strategy for the heterogeneous platoon over the S-Road profile. The front truck $(1)$ is lighter loaded and so drops its velocity uphill to accommodate the heavier following truck $(2)$.}
    \label{fig:Constraint}
\end{figure}

Figure \ref{fig:Constraint} then examines the performance of the trucks $k=1$ and $k=2$ for the considerate strategy. Here it can be seen that the leading truck throttles down its speed during the uphill portion as a necessary method to maintain a close gap with the truck behind it, and thereby behaves similar to as if it were a heavier truck. As in the motivating experimental scenario shown in Figure \ref{fig:Motivation}, the trailing truck $k=2$ is seen here to maximize its engine to its capabilities during the uphill portion, and so is at the limit of its climbing ability and requires the assistance of $k=1$ to remain near its target gap.

The 3-truck platoon is then simulated on the \unit[70]{km} stretch of U.S. highway travelling from Lanesville, IN to Siberia, IN for a nominal model of road grade conditions for heavy-duty long-haul truck applications \cite{Liu2019}. A batch of scenarios is generated where each truck is loaded with a mass sampled from $m \in \{14, 22, 30, 38\}$t and all possible permutations of truck orderings with these loadings are simulated (24 total simulations). As before, the considerate and anticipative MPC variants are used to guide the platoon through the route.

\begin{figure}
    \centering
    \input{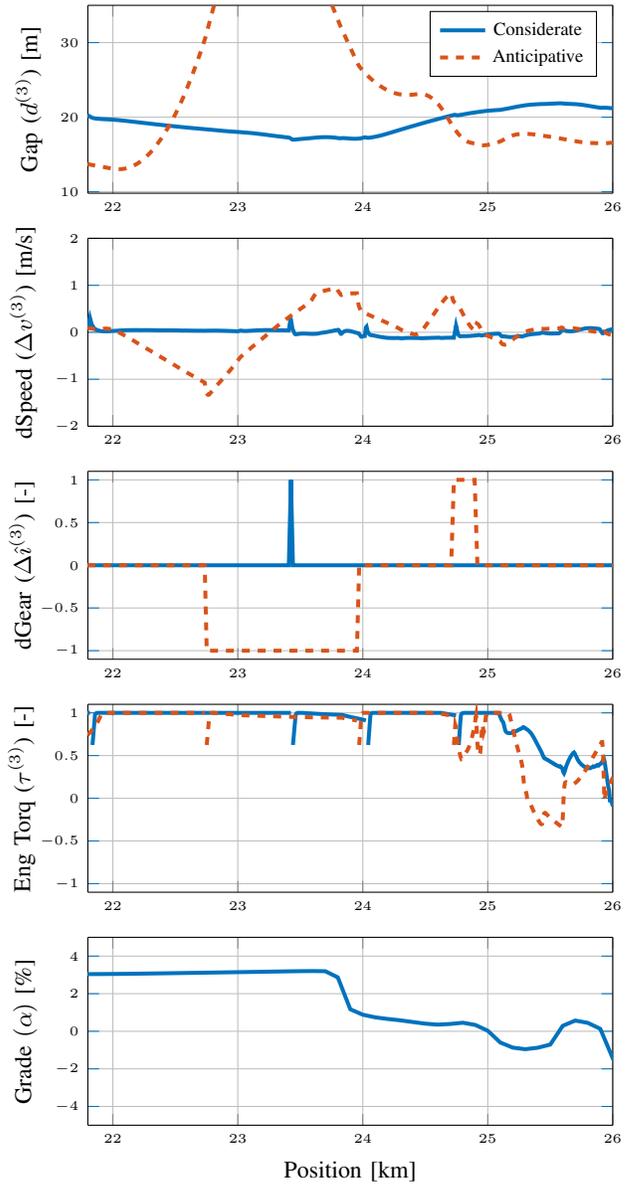}
    \caption{Excerpt of Lanesville-Siberia route \cite{Liu2019}. Considerate MPC (blue, solid) improved harmonization between the trucks and reduced transients compared to the anticipative MPC (red, dashed). Operator $\Delta(\cdot)$ indicates difference in ego truck quantity subtracted by preceding truck quantity.}
    \label{fig:Lanesville}
\end{figure}

Figure \ref{fig:Lanesville} depicts an excerpt of the route for the final truck in the platoon when loaded with the first-to-last configuration $\{30, 38, 38\}$t. It is seen that the considerate control improved velocity harmonization between trucks, and as a result better improved the harmonization between selected gears as determined by the automatic transmission. Gap tracking performance was similarly steadied in contrast to the anticipative strategy during the uphill portion of the road segment, despite that the engine torque is maximized.

Finally, the mean results - with one standard deviation denoted as $(\cdot)$ - of the batch simulations on the route are summarized in Tables \ref{tab:summary_c} and \ref{tab:summary_nc} for the considerate and anticipative variants, respectively. Travel time and distance travelled are the same in all cases. For this route, the A-MPC strategy still performed well in terms of fuel performance and is a known improvement over classical CACC due to reduced engine demand \cite{ARD2020}, but similarly to the CACC shown in Section \ref{sec:Motivation}, exhibits large velocity transients, poor harmonization, and is prone to disengagement that requires operator intervention in nearly half of the scenarios. Overall, the C-MPC had greatly consistent performance with its low standard deviations in each metric and significant improvements to root-mean-square-error (RMSE) in target gap tracking from the preceding truck, as well as a moderate 1.8\% improvement in fuel economy for the entire platoon. 

    
\begin{table}[]
    \centering
    \caption{Performance of the considerate MPC.}
    \begin{tabular}{c|c|c|c|c}
         \toprule
         $k$ & Fuel [kg/100km] & Headway [s] & Gap \scriptsize{RMSE} [m] & Diseng. [-] \\
         \midrule
$1$ &	$5.16 \ (0.29)$ &	- &	- & - \\
$2$ &	$4.90 \ (0.28)$ &	$0.72 \ (0.01)$ &	$1.25 \ (0.17)$ & 0 \\
$3$ &	$4.87 \ (0.25)$ &	$0.68 \ (0.01)$ &	$4.60 \ (0.25)$ & 0 \\
        \bottomrule
    \end{tabular}
    \label{tab:summary_c}
\end{table}
\begin{table}[]
    \centering
    \caption{Performance of the anticipative MPC.}
    \begin{tabular}{c|c|c|c|c}
         \toprule
         $k$ & Fuel [kg/100km] & Headway [s] & Gap \scriptsize{RMSE} [m] & Diseng. [-] \\
         \midrule
$1$ &	$5.12 \ (0.26)$ &	- &	- & - \\
$2$ &	$5.03 \ (0.41)$ &	$1.40 \ (0.14)$ &	$276.7 \ (43.1)$ & $0.50 \ (0.5)$ \\
$3$ &	$5.05 \ (0.38)$ &	$0.94 \ (0.41)$ &	$36.09 \ (46.5)$ & $0.38 \ (0.5)$\\
        \bottomrule
    \end{tabular}
    \label{tab:summary_nc}
\end{table}

\section{Conclusion}
This paper developed a distributed model predictive control that is considerate of interacting agents to address experimentally evaluated limitations in heavy-duty truck platooning if deploying a classically designed CACC, in which we propose that assistance is needed from leader trucks in the platoon to maintain stability under the presence of disturbances and actuator limitations. The MPC forms an optimization of the ego truck that is considerate of its follower to determine a sequence of actions it can take that benefit both trucks. The considerate controller was then tasked to drive a multi-truck platoon through a nominal-shaped corridor and a corridor based on a U.S highway, and its performance was compared against a non-considerate variant of the controller. Overall, it was found that the considerate strategy significantly improved harmonization between the platooned trucks without harming real-time feasibility.


\section*{Acknowledgment}
 
 This material is based upon work supported by the Department of Energy, Office of Energy Efficiency and Renewable Energy (EERE), under Award Number DE-EE0008469. 
\bibliographystyle{IEEEtran}
\bibliography{accbib}

\begin{thebibliography}{10}
\providecommand{\url}[1]{#1}
\csname url@samestyle\endcsname
\providecommand{\newblock}{\relax}
\providecommand{\bibinfo}[2]{#2}
\providecommand{\BIBentrySTDinterwordspacing}{\spaceskip=0pt\relax}
\providecommand{\BIBentryALTinterwordstretchfactor}{4}
\providecommand{\BIBentryALTinterwordspacing}{\spaceskip=\fontdimen2\font plus
\BIBentryALTinterwordstretchfactor\fontdimen3\font minus
  \fontdimen4\font\relax}
\providecommand{\BIBforeignlanguage}[2]{{%
\expandafter\ifx\csname l@#1\endcsname\relax
\typeout{** WARNING: IEEEtran.bst: No hyphenation pattern has been}%
\typeout{** loaded for the language `#1'. Using the pattern for}%
\typeout{** the default language instead.}%
\else
\language=\csname l@#1\endcsname
\fi
#2}}
\providecommand{\BIBdecl}{\relax}
\BIBdecl

\bibitem{Shladover2012}
S.~E. Shladover, D.~Su, and X.~Y. Lu, ``{Impacts of cooperative adaptive cruise
  control on freeway traffic flow},'' \emph{Transportation Research Record},
  vol. 2324, pp. 63--70, 2012.

\bibitem{Liu2020}
H.~Liu, S.~E. Shladover, X.~Y. Lu, and X.~Kan, ``{Freeway vehicle fuel
  efficiency improvement via cooperative adaptive cruise control},''
  \emph{Journal of Intelligent Transportation Systems: Technology, Planning,
  and Operations}, 2020.

\bibitem{Wang2020}
Z.~Wang, Y.~Bian, S.~E. Shladover, G.~Wu, S.~E. Li, and M.~J. Barth, ``{A
  Survey on Cooperative Longitudinal Motion Control of Multiple Connected and
  Automated Vehicles},'' \emph{IEEE Intelligent Transportation Systems
  Magazine}, vol.~12, no.~1, pp. 4--24, 2020.

\bibitem{McAuliffe2018}
B.~McAuliffe, M.~Lammert, X.~Y. Lu, S.~Shladover, M.~D. Surcel, and A.~Kailas,
  ``{Influences on Energy Savings of Heavy Trucks Using Cooperative Adaptive
  Cruise Control},'' \emph{SAE Technical Papers}, vol. 2018-April, no. April,
  pp. 10--12, 2018.

\bibitem{Borhan2021}
H.~Borhan, M.~Lammert, K.~Kelly, C.~Zhang, N.~Brady, C.~S. Yu, and J.~Liu,
  ``{Advancing Platooning with ADAS Control Integration and Assessment Test
  Results},'' \emph{SAE Technical Papers}, no. 2021, pp. 1969--1975, 2021.

\bibitem{Kerrigan2000}
E.~C. Kerrigan and J.~M. Maciejowski, ``{Invariant sets for constrained
  nonlinear discrete-time systems with application to feasibility in model
  predictive control},'' \emph{Proceedings of the IEEE Conference on Decision
  and Control}, vol.~5, pp. 4951--4956, 2000.

\bibitem{Keviczky2006}
T.~Keviczky, F.~Borrelli, and G.~J. Balas, ``{Decentralized receding horizon
  control for large scale dynamically decoupled systems},'' \emph{Automatica},
  vol.~42, no.~12, pp. 2105--2115, 2006.

\bibitem{Dunbar2006}
W.~B. Dunbar and R.~M. Murray, ``{Distributed receding horizon control for
  multi-vehicle formation stabilization},'' \emph{Automatica}, vol.~42, no.~4,
  pp. 549--558, 2006.

\bibitem{Zheng2017}
Y.~Zheng, S.~E. Li, K.~Li, F.~Borrelli, and J.~K. Hedrick, ``{Distributed Model
  Predictive Control for Heterogeneous Vehicle Platoons under Unidirectional
  Topologies},'' \emph{IEEE Transactions on Control Systems Technology},
  vol.~25, no.~3, pp. 899--910, 2017.

\bibitem{Wang2015}
J.~Q. Wang, S.~E. Li, Y.~Zheng, and X.~Y. Lu, ``{Longitudinal collision
  mitigation via coordinated braking of multiple vehicles using model
  predictive control},'' \emph{Integrated Computer-Aided Engineering}, vol.~22,
  no.~2, pp. 171--185, 2015.

\bibitem{Turri2017}
V.~Turri, Y.~Kim, J.~Guanetti, K.~H. Johansson, and F.~Borrelli, ``{A model
  predictive controller for non-cooperative eco-platooning},''
  \emph{Proceedings of the American Control Conference}, no. iii, pp.
  2309--2314, 2017.

\bibitem{Turri2017b}
V.~Turri, B.~Besselink, and K.~H. Johansson, ``{Cooperative Look-Ahead Control
  for Fuel-Efficient and Safe Heavy-Duty Vehicle Platooning},'' \emph{IEEE
  Transactions on Control Systems Technology}, vol.~25, no.~1, pp. 12--28,
  2017.

\bibitem{Ibrahim2019}
A.~Ibrahim, M.~Cicic, D.~Goswami, T.~Basten, and K.~H. Johansson, ``{Control of
  Platooned Vehicles in Presence of Traffic Shock Waves},'' \emph{2019 IEEE
  Intelligent Transportation Systems Conference, ITSC 2019}, pp. 1727--1734,
  2019.

\bibitem{He2016}
C.~R. He, H.~Maurer, and G.~Orosz, ``{Fuel Consumption Optimization of
  Heavy-Duty Vehicles with Grade, Wind, and Traffic Information},''
  \emph{Journal of Computational and Nonlinear Dynamics}, vol.~11, no.~6, pp.
  1--12, 2016.

\bibitem{Hellstrom2009}
E.~Hellstr{\"{o}}m, M.~Ivarsson, J.~{\AA}slund, and L.~Nielsen, ``{Look-ahead
  control for heavy trucks to minimize trip time and fuel consumption},''
  \emph{Control Engineering Practice}, vol.~17, no.~2, pp. 245--254, 2009.

\bibitem{Ploeg2011}
J.~Ploeg, B.~T. Scheepers, E.~{Van Nunen}, N.~{Van De Wouw}, and H.~Nijmeijer,
  ``{Design and experimental evaluation of cooperative adaptive cruise
  control},'' \emph{IEEE Conference on Intelligent Transportation Systems,
  Proceedings, ITSC}, pp. 260--265, 2011.

\bibitem{GUZZELLA2013}
L.~Guzzella and A.~Sciarretta, \emph{Vehicle Propulsion Systems}, 3rd~ed.\hskip
  1em plus 0.5em minus 0.4em\relax Springer, 2013.

\bibitem{ARD2020}
T.~Ard, F.~Ashtiani, A.~Vahidi, and H.~Borhan, ``Optimizing gap tracking
  subject to dynamic losses via connected and anticipative {MPC} in truck
  platooning,'' in \emph{2020 American Control Conference (ACC)}, 2020, pp.
  2300--2305.

\bibitem{FORCESNLP}
A.~Zanelli, A.~Domahidi, J.~Jerez, and M.~Morari, ``Forces nlp: an efficient
  implementation of interior-point methods for multistage nonlinear nonconvex
  programs,'' \emph{International Journal of Control}, pp. 1--17, 2017.

\bibitem{Andersson2019}
J.~A.~E. Andersson, J.~Gillis, G.~Horn, J.~B. Rawlings, and M.~Diehl,
  ``{CasADi} -- {A} software framework for nonlinear optimization and optimal
  control,'' \emph{Mathematical Programming Computation}, vol.~11, no.~1, pp.
  1--36, 2019.

\bibitem{Liu2019}
J.~Liu, B.~Pattel, A.~S. Desai, E.~Hodzen, and H.~Borhan, ``{Fuel Efficient
  Control Algorithms for Connected and Automated Line-Haul Trucks},''
  \emph{CCTA 2019 - 3rd IEEE Conference on Control Technology and
  Applications}, pp. 730--737, 2019.

\end{thebibliography}

\end{document}